\documentstyle[multicol,prd,aps,graphicx]{revtex}

\begin{document}
\vspace{.1in}

\draft
\title{ 
\vglue -0.5cm
\vglue 0.5cm
Neutrino masses through see-saw mechanism in 3-3-1
models} 
\author{\bf J. C. Montero\footnote{montero@ift.unesp.br}, C. A. de S. 
Pires\footnote{cpires@ift.unesp.br} and
V. Pleitez\footnote{ vicente@ift.unesp.br} }
\address{
Instituto de F\'\i sica Te\'orica\\
Universidade Estadual Paulista\\
Rua Pamplona, 145\\ 
01405-900-- S\~ao Paulo, SP\\
Brazil}
\date{\today}

\maketitle
\begin{abstract}
Some years ago it was shown by Ma that in the context of the electroweak 
standard model there are, at the tree level, only three ways 
to generate small neutrino masses by the see-saw mechanism 
via one effective dimension-five operator. 
Here we extend this approach to 3-3-1 chiral models showing that in this case
there are several dimension-five operators and we also consider their tree
level realization. 
\end{abstract}



\begin{multicols}{2}
\narrowtext

\section{Introduction}
\label{sec:intro}

Although recent data on neutrino oscillation experiments~\cite{nussol,nusat,lsnd}
strongly suggest that neutrinos have nonzero small masses the issue of
explaining its tiny value is still an open question.
Several years ago it was noted by Weinberg~\cite{sw} and independently by
Wilczek and Zee~\cite{wz}, that the neutrinos may acquire naturally small
Majorana masses through dimension-five effective operators like 
\begin{equation}
\overline{l^c_{iaL}}\,l_{jbL}\Phi^{(m)}_k\Phi^{(n)}_l
(f_{abmn} \epsilon_{ik}\epsilon_{jl}+
f^\prime_{abmn} \epsilon_{ij}\epsilon_{jl}),
\label{e1}
\end{equation}
being the couplings $f$ and 
$f^\prime$ of the order of $\Lambda^{-1}$, where $\Lambda$ is a large effective
mass related to new physics. When the neutral component 
of the scalar doublet $\Phi$ develops its vacuum expectation value (VEV), 
$\langle \phi \rangle $, it will produce the following mass matrix for the 
neutrinos 
\begin{equation}
M_\nu = \frac{f \langle \phi \rangle ^2 }{\Lambda},
\label{seesaw}
\end{equation}
and since $ \langle \phi \rangle$ is of the order of $100$
GeV, small Majorana neutrino masses are generated if $\Lambda > 10^{13}$ GeV.
The interesting point in Eq.~(\ref{seesaw}) is that the lightness of the
neutrino masses are generated via new physics at an energy scale $\Lambda\gg
G^{-1/2}_F$. This is the usual see-saw mechanism~\cite{seesaw}. 
The realizations of that operator inside the standard model (SM) were already 
investigated by Ma in Ref.~\cite{ema}. There it was sketched all the ways such
operator can  be realized at the  tree and one loop level. 

Here we re-derive step by step such realizations and extend the analysis to 
3-3-1 chiral models~\cite{331,lepmass1,bbd331} in which, like in the 
SM, neutrinos are massless unless we add right-handed neutrinos
or break the lepton number. In this vain, after SuperKamiokande 
results, several suitable modifications of the model were already 
proposed in order to generate the neutrino masses~\cite{phf1,yasue}.

In this work we will build in the framework of the  3-3-1 models an effective 
dimension-five operator that leads to see-saw neutrino masses and then  
investigate how it can be realized at the tree and one loop level. 
Although it was noted in the early references~\cite{sw,wz} that it is necessary
to renormalize these effective operators, we will not address this issue
here~\cite{racorr}. 

The outline of this work is the following. In section II we review the 
several ways to generate such sort of operators in the context of the 
standard model. In section III we consider the generation of those operators 
in the 3-3-1 model. Our conclusion will appear in section IV.  

\section{See-saw tree level realization in the standard model}
\label{sec:sm}

In the context of the standard model with only the lepton doublets
$\Psi_{aL}=(\nu_{aL},l_{aL})^T,\;a=e,\mu,\tau$, and a scalar doublet
$\Phi=(\phi^+,\phi^0)^T$, there are only three possibilities for implementing
naturally small neutrino masses with dimension-five effective
operators~\cite{ema}. On the other hand, if we add new leptons a new possibility
arises. They are the following:

{\bf I}) $\Psi_{aL}$ and $\Phi$ form a fermion singlet, so that the
effective interaction is
\begin{eqnarray}
{\cal L}^{eff}_{1F}&=&\frac{f_{ab}}{\Lambda}[\epsilon
\overline{(\Psi^c)_{aR}}\,\Phi][\epsilon
\overline{(\Psi^c)_{bR}}\,\Phi]+H.c.\nonumber \\ &=&
\frac{f_{ab}}{\Lambda}\left(\overline{(\nu^c)_{aR}}\,\phi^0-
\overline{(l^c)_{aR}}\,\phi^+ \right)\left( \nu_{bL}\phi^0-l_{bL}\phi^+\right)
\nonumber \\ &+& H.c.,
\label{eff1}
\end{eqnarray}
$\epsilon$ denotes the antisymmetric $SU(2)$ tensor and we have suppressed
$SU(2)$ indices. The tree level realization of this effective 
operator is achieved by the introduction of one or more neutral right-handed 
singlets, say $N_{bR}$, which produces the usual see-saw 
mechanism~\cite{seesaw}, and it is linked to the standard left-handed 
neutrinos through the interaction
\begin{equation}
f_{ab}\bar \Psi_{aL} \tilde\Phi N_{bR} +H.c.,
\label{diracmass}
\end{equation}
with $\tilde\Phi=i\tau^2\Phi^*$, this interaction and the bare mass term
for $N_R$, $\overline{(N_{aR})^c}  M_{Rab} N_{bR}$, leads to the following
Dirac-Majorana matrix 
\begin{equation}
\left(
\begin{array}{cc}
0 & M^T_D\\ M_D & M_R\end{array}
\right),
\label{D-Mmatrix}
\end{equation}
in the basis $(\nu_L \, , \, (N_R)^c)^T $. The matrix above, after 
diagonalization, gives the following expression for the neutrino masses:
\begin{equation}
M_\nu = M^T_D M^{-1}_R M_D.
\label{seesawI}
\end{equation}

\begin{figure}
\begin{center}
\includegraphics[width=5cm]{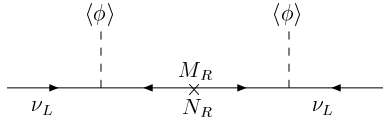}
\end{center}
\caption{
Tree level realization of the effective 
dimension-five operator through a heavy neutrino.
}
\label{fig1}
\end{figure}

From Eq.~(\ref{seesawI}) we recover Eq.~(\ref{seesaw}) by choosing 
$M_D = f\langle \phi \rangle$ and taking $M_R = \Lambda$, obtaining in this way 
the canonical see-saw mechanism, or the now called type I~\cite{seesaw}, which 
is the usual realization of the effective operator in Eq.~(\ref{e1}) found 
in the literature. Notice that the see-saw here is a relation among fermion masses and 
$N_R$ could be the right-handed neutrinos $\nu_R$. 
Its realization is depicted in Fig.~\ref{fig1} with  the replacements
$N_R\to \nu_R $ and $ \langle \Phi \rangle \to \langle \phi \rangle $.

{\bf II}) $\Psi_{aL}$ and $\Psi_{bL}$ form a scalar triplet with the effective
interaction
\begin{eqnarray}
{\cal L}^{eff}_{3S}&=&\frac{f_{ab}}{\Lambda}\,(\overline{\Psi^c)_{aR}}\Psi_{bL}
(\epsilon\Phi\cdot\epsilon\Phi)+H.c. \nonumber \\ &=&
\frac{f_{ab}}{\Lambda}\,[\overline{(\nu^c)_{aR}}\nu_{bL}\phi^0\phi^0-
2\phi^0\phi^+(\bar\nu^c_{aR}l_{bL}
+\bar{l}^c_{aR}\nu_{bL})\nonumber \\ &+&\bar{l}^c_{aR}l_{bL}\phi^+\phi^+]
+H.c..
\label{eff2}
\end{eqnarray}
The tree level realization in this case is obtained by
introducing a complex ($Y=2$) scalar triplet 
$(\xi^{++},\xi^+,\xi^0)$ or, in matrix notation~\cite{triplet,gr},
\begin{equation}
\vec\tau\cdot\vec\Xi=
\left(\begin{array}{cc}
\xi^+&\sqrt{2}\xi^{++}\\
\sqrt{2}\xi^0&-\xi^+\end{array}\right).
\label{ii}
\end{equation}

In this case we have the interaction between the triplet $\Xi$ and
the usual leptons,
\begin{equation}
 f_{ab}\bar \Psi^c_{aL} (\epsilon\vec{\tau}\cdot\Xi) \Psi_{bL} + H.c.
\label{inttrip}
\end{equation}

From the expression above $\Xi$ must carry two units of lepton number. 
After the triplet develops a VEV the neutrinos gain masses which are given 
by the following expression:
\begin{equation}
M_{\nu ab}=f_{ab}\langle \xi^0 \rangle.
\label{netrmaII}
\end{equation}

Differently from the type I see-saw mechanism here the suppression must come 
from $\langle\xi^0\rangle(\equiv\langle \xi \rangle)$. For this we  have to 
study the scalar potential of the model. The complete scalar potential 
composed by the standard Higgs doublet $\Phi$ and the triplet $\Xi$ is: 
\begin{eqnarray}
V(\phi,\xi) &=& \mu^2_\phi \Phi^{\dagger} \Phi + \mu_\Xi^2 
\Xi^{\dagger} \Xi  
+\lambda_\phi (\Phi^{\dagger} \Phi)^2  + \lambda_\Xi(\Xi^{\dagger} 
\Xi)^2 
\nonumber \\
&&\lambda_{\Xi \phi}\Xi^{\dagger} \Xi \Phi^{\dagger} \Phi + 
M_\Xi \Phi^T \Xi^{\dagger} \Phi ,
\label{grp} 
\end{eqnarray}
where the last term breaks explicitly the lepton number. The stationary
conditions for this potential gives the following constraint equations: 
\begin{eqnarray}
&&\langle \phi \rangle (\mu_\phi^2 +\lambda_\phi \langle \phi 
\rangle^2 
+ 
\lambda_{\xi \phi}\langle \xi \rangle^2)=0,\nonumber \\
&&\langle \xi \rangle( \mu_\xi^2 +\lambda_\xi \langle \xi \rangle^2 
+\lambda_{\xi \phi} \langle \phi \rangle^2 ) + M_\xi 
\langle \phi \rangle^2=0.
\label{tadpole}
\end{eqnarray}
Assuming that $\vert\mu_\xi \vert\sim M_\xi \gg \langle \phi \rangle$, with
$\mu^2_\xi<0$,  the 
second constraint equations provides the following relation between the vacua 
of the model:
\begin{equation}
\langle \xi \rangle \sim \frac{\langle \phi \rangle^2}{M_\xi}.
\label{tinyd}
\end{equation}

Substituting Eq.~(\ref{tinyd}) in Eq.~(\ref{netrmaII}) we get the see-saw  mass 
relation for neutrinos,
\begin{equation}
M_{\nu ab} = \frac{f_{ab} \langle \phi \rangle^2}{M_\xi}.
\label{typeIIss}
\end{equation}

\begin{figure}
\begin{center}
\includegraphics[width=3cm]{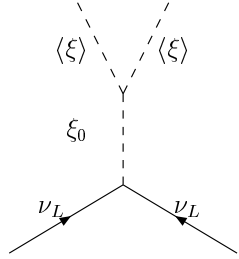}
\end{center}
\caption{
Tree level realization of the effective 
dimension-five operator through a heavy 
scalar.
}
\label{fig2}
\end{figure}

If the triplet $\xi$ belongs to some GUT, we can recognize $M_\xi$ as the high
scale $\Lambda$. This is the so called type II see-saw
mechanism~\cite{seesawII}. Notice that in  this type of see-saw a very heavy
scalar could develop a very tiny vacuum~\cite{ema,typeII}, differently of the
type I see-saw where a very heavy right-handed neutrino  induces tiny mass to the
light neutrinos. The tree level realization of this mechanism is shown in
Fig.~\ref{fig2}. 

{\bf III}) $\Psi_{aL}$ and $\Phi$ form a fermion triplet with $Y=0$ and 
the effective Lagrangian is then written as:
\begin{eqnarray}
{\cal L}^{eff}_{3F}&=&\frac{f_{ab}}{\Lambda}\,[\overline{(\Psi^c)_{aR}}
\cdot\Phi](\epsilon\Psi_{bL}\cdot\epsilon\Phi)+H.c.\nonumber \\ &=&
\frac{f_{ab}}{\Lambda}\,[\bar\nu^c_{aR}l_{bL}\phi^+\phi^0-(\bar\nu^c_{aR}\phi^0+
\bar{l}^c_{aR}\phi^+)\nonumber \\ &&\mbox{}
(l_{bL}\phi^++\nu_{bL}\phi^0)+
\bar{l}^c_{aR}\nu_{bL}\phi^+\phi^0]+H.c..
\label{eff3}
\end{eqnarray}

In this case, the tree level realization is based on the introduction of the 
$Y=0$ triplet of leptons~\cite{rf}:
\begin{equation}
\vec{\tau}\cdot\vec{\Delta}_L=\left(
\begin{array}{cc}
N & \sqrt{2}P^+\\ \sqrt{2}P^-& -N\end{array}
\right)_L.
\label{ftriplet}
\end{equation}

This triplet is linked to the standard model content by the interaction
\begin{equation}
f_a\,\left[\, \tilde\Phi^T\overline{ \vec{\tau}\cdot\vec\Delta_L}\, 
(\Psi_{aL})^c + \Phi^T\,\overline{(\vec{\tau}\cdot\Delta_L)^c}\, \Psi_{aL} 
\right] +H.c.,
\label{deltaint}
\end{equation}
where $\tilde\Phi=\epsilon\Phi^*$.
The interaction in Eq.~(\ref{deltaint}),
together with the mass term  $M_\Delta\overline{(\Delta_L)^c}\Delta_L$, 
yields the mass matrix, in the basis $(\nu_{aL}, \, N_L)^T$, 
\begin{equation}
\left(
\begin{array}{cc}
0 & M_L \\ M_L & M_\Delta 
\end{array}
\right).
\label{masstriplet}
\end{equation}

Notice that in this case $M_{aL}=f_a\langle\phi\rangle$ is not an arbitrary
mass matrix, hence, after diagonalization, two neutrinos are massless and two 
have non-zero masses at the tree level. One of the massive neutrinos has a mass
proportional to $M_\Delta$ and the other one $\sim
f_a\langle\phi\rangle^2/M_\Delta$~\cite{rf}.  
Although this is also a see-saw  relation we see that the neutrino mass spectrum
is different from that of the general see-saw mechanism, since two of them are 
massless at the tree level. Notice also that the interactions in
Eq.~(\ref{deltaint}) violate explicitly the lepton number. This is the third
possibility of realization at the tree level of the effective dimension-five
operator that leads to a see-saw relation  for the neutrino masses. Even though
it leads to the same see-saw relation, as the last two most explored mechanisms
explicited in {\bf I}) and {\bf II}), this mechanism has not been widely
appreciated in literature. It is depicted in Fig.~\ref{fig1} with $N_R $
replaced by $(N_L)^c $. 

An interesting version of the model in which we obtain a general see-saw mass
matrix is the one where we identify $N_L$ as being
$(\nu^c)_L$ and add three triplets $\Delta_{aL},\,a=e,\mu,\tau$; since in this
case $M_L=f_{ab}\langle\phi\rangle$, we can have a neutrino mass matrix as in
Eq.~(\ref{seesawI}) in the canonical see-saw mechanism. 

Unlike the other two types of see-saw mechanisms, in the present one we must
verify if this fermion triplet $\Delta$ does not lead to any  implication in
the charged lepton masses. 
The general mass matrix in the charged lepton
sector is rather complicated since the interactions in
Eq.~(\ref{deltaint}) together with the standard Yukawa interaction $\lambda \bar
\Psi_L \Phi l_R$ lead to the following charged lepton matrix in the basis 
$(l_L,\,\,P_L)$:
\begin{equation}
( \bar l_L \,\,\, \bar P_L ) \left(
\begin{array}{cc}
M_D & M_L \\ 0 & M_\Delta\end{array}
\right)\left(
\begin{array}{cc}
l_R  \\
P_R\end{array}
\right),
\label{leptonmass}
\end{equation}
where $M_{Dab} = \lambda_{ab} \langle \phi \rangle$.
With $M_{aL} = f_a\langle 
\phi \rangle$ (or $M_L=f_{ab}\langle\phi\rangle$ if we add three triplets) 
we see that the mass matrix in
Eq.~(\ref{leptonmass}) can be only easily diagonalized if we assume that $M_D$
is diagonal. For more details see Ref.~\cite{rf}. 

{\bf IV)} However, if we extent the particle content of the standard model,
another possibility is that $\Psi^c_{aR}$ and $\Phi$ form a 
non-Hermitian fermionic triplet with $Y=2$. In this case, although it is
not possible to built an effective operator of the form given in 
Eq.~(\ref{eff1}), we can introduce a fermionic non-Hermitian ($Y=2$) triplet
$\Omega=(\omega^{++},\omega^+,\omega^0)$  or 
\begin{equation}
\vec\tau\cdot\vec\Omega=
\left(\begin{array}{cc}
\omega^+&\sqrt{2}\omega^{++}\\
\sqrt{2}\omega^0&-\omega^+\end{array}\right)_L,
\label{iv}
\end{equation}
so that we have the interactions
\begin{equation}
f_a\,\left[\, \Phi^T\overline{ \vec{\tau}\cdot\vec\Omega_L}\, 
(\Psi_{aL})^c + \tilde\Phi^T\,\overline{(\vec{\tau}\cdot\Omega_L)^c}
\, \Psi_{aL}\right] +H.c.,
\label{omegaint}
\end{equation}
which together with a mass term $M_\Omega \overline{(\Omega^c)_R}\Omega_L$ give 
a mass matrix of the form in Eq.~(\ref{D-Mmatrix}) with 
$M_R\to M_\Omega$. 
Next, we introduce the right-handed doublets as
$\psi_{aR}=(N,E^-)_{aR}\sim({\bf2},-1)$, and so it is 
possible to have 
\begin{equation}
{\cal L}^{eff}_{4F}=\frac{f_{ab}}{\Lambda}\,\left(\overline{\Psi_{aL}}
\cdot\Phi\right)\left(\psi_{bR}\cdot\tilde\Phi\right)+H.c.
\label{eff4}
\end{equation}

Notice that, even if the first three ways  of realization of the effective 
dimension-five operator at tree level are very different, in the end the 
neutrino masses receive the same expression in the form of the
see-saw relation presented in Eq.~(\ref{seesaw}). 
In the fourth way it is also
possible to generate such a mass matrix but only at the tree level if we do not
restrict ourselves to the usual leptonic representation content.
The only context in which they can be distinguished is by their
consequences at high energy once the type I see-saw favors $SO(10)$ and 
type II and III favor $SU(5)$ in {\bf 15}  and {\bf 24} representation 
respectively. In fact, the first one has been more studied in the 
literature and it was in this context
that the see-saw mechanism was proposed~\cite{seesaw}. The scalar triplet has
also been largely consider in literature~\cite{triplet}. Both alternatives can
be implemented more naturally in the context of the standard model or in some of
its extensions since, for instance, the introduction of neutral singlets leaves
the model more symmetric relatively to quarks and leptons, or because the
scalar representation content is not constrained by the gauge invariance. This
is not the case with the lepton triplet. However, recently it has been shown 
that in a 3-3-1 supersymmetric model~\cite{lepmass2} there is a fermionic
non-hermitian triplet under the 3-2-1 symmetry that is part of a sextet of 
higgssinos transforming as $({\bf6},0)$ under $SU(3)_L\otimes U(1)_N$. 

\section{ the effective dimension-five operator and its realization 
in 3-3-1 chiral models}
\label{sec:331}

In this section we build the  effective dimension-five operator in the
context of 3-3-1 chiral models and its realizations at the tree and the 1-loop 
level. 
We employ the same approach used for the realizations of Eq.~(\ref{seesaw}) in
the SM outlined in the last section. As in the SM, in the minimal 3-3-1
model~\cite{331}, or in some of its extensions~\cite{yasue}, neutrinos are
strictly massless due to the conservation of the total lepton number. In
these models all leptons come only in triplets. In one version of this model,
to generate the mass of all the particle content of the model, three triplets
$\eta$,  $\rho$, $\chi$ and a sextet $S$ of scalars are required.  

We have at least two triplets, the leptonic one
$\Psi_{aL}=(\nu_{aL}, l_{aL},l^c_{aL})^T$ and the
scalar one $\eta=(\eta^0,\eta^-_1,\eta^+_2)^T$, both transforming like
$({\bf3},0)$ under the electroweak gauge symmetry $SU(3)_L\otimes U(1)_N$.
Hence, we can form bilinears like $\bar\Psi\eta=\bar{\bf
3}\otimes{\bf3}={\bf1}_A\oplus {\bf8}$ or 
$\Psi\eta={\bf3}\otimes{\bf3}=\bar{\bf3}_A\oplus{\bf6}_S$. In this case we will
have the following possibilities for the effective 
interactions:

{\bf A}) $\Psi$ and $\eta$ form a fermion singlet, and the effective operator is
written as 
\begin{eqnarray}
{\cal L}^{eff}_{1F}&=&\frac{f_{ab}}{\Lambda}
[\eta^\dagger\overline{(\Psi^c)_{aR}}][\eta^\dagger \Psi_{bL}]+H.c.
\nonumber \\ &=&\frac{f_{ab}}{\Lambda}(\bar\nu^c_{aR}\eta^{0*}+\bar{l}^c_{aR}
\eta^+_1+\bar{l}_{aR}\eta^-_2)
\nonumber \\ &&\mbox{}(\nu_{bL}\eta^{0*}+l_{bL}\eta^+_1+l^c_{bL}\eta^-_2)+H.c.
\label{331a}
\end{eqnarray}

At the tree level we can realize this situation by introducing again one or more
neutral fermion singlets, $N_{bR}$'s, and it is also a realization of the usual 
see-saw mechanism. This is in fact equivalent to the case {\bf I}) in the 
previous section. Like  there the intermediate heavy particle here should be a
right-handed neutrino,  $\nu_R$. The realization of this operator with $\nu_R$ 
is similar to that  in Fig.~\ref{fig1} with $\langle \phi \rangle$ replaced by
$\langle \eta \rangle$. 
The heavy right-handed neutrinos are linked with $\nu_L$ by a term
like this one:
\begin{equation}
 f_{ab} \bar \Psi_{aL}\eta \,\nu_{bR} + H.c,
\label{linkint}
\end{equation}
which together with the mass term $M_R\bar \nu^c_R \nu_R$ leads to the 
mass matrix defined in Eq.~(\ref{D-Mmatrix}) in the basis $(\nu_L, \,
(\nu_R)^c)$ with $M_{Dab} = f_{ab}\langle\eta \rangle $.  

{\bf B}) Next, we can form with $\Psi_{aL}$ and $\eta$ an octet of leptons if we
define the traceless matrix,
\begin{equation}
\bar{M}^j_{ai}=\overline{(\Psi^c)_{aiR}}\,\eta^{\dagger j}-\frac{1}{3}\,
\delta^j_i\,(\overline{(\Psi^c)_{aR}}\cdot\eta^\dagger), 
\label{m}
\end{equation}
where $i,j$ denote $SU(3)$ indices. The effective interaction is in this case 
given by
\end{multicols}
\hspace{-0.5cm}
\rule{8.7cm}{0.1mm}\rule{0.1mm}{2mm}
\widetext
\begin{eqnarray}
{\cal L}^{eff}_{8F}&=&\frac{f_{ab}}{\Lambda}\;{\rm Tr}(\bar{M}_aM_b)+H.c.
\nonumber \\ &=&
\frac{f_{ab}}{\Lambda}\,[\frac{1}{9}(2\bar\nu^c_{qR}\eta^{0*}-
\bar{l}^c_{aR}\eta^+_1-\bar{l}_{aR}\eta^-_2)
(2\nu_{bL}\eta^{0}-l^c_{bL}\eta^-_1-l_{bL}\eta^+_2)+
\bar\nu^c_{aR}\nu_{b:}(\eta^+_1\eta^-_1+ \eta^+_2\eta^-_2)
\nonumber \\
&+&\bar{l}^c_{aR}l_{bL}(\eta^{0*}\eta^0+\eta^+_2\eta^-_2) 
+\frac{1}{9}(2\bar{l}^c_{aR}\eta^+_1-\bar\nu^c_{aR}\eta^{0*}-
\bar{l}_{aR}\eta^-_2)(2l^c_{bL}\eta^-_1-\nu_{bL}\eta^0-l_{bL}\eta^+_2)+
\bar{l}^c_{aR}l^c_{bL}\eta^-_2\eta^+_1\nonumber \\ &+&
\bar{l}_{aR}l_{bL}(\eta^{0*}\eta^0+\eta^+_1\eta^-_1)
+\frac{1}{9}(2\bar{l}_{aR}\eta^-_2-\bar{l}^c_{aR}
\eta^+_1-\bar{\nu}^c_{aR}\eta^{0*})(2l_{bL}\eta^+_2-l^c_{bL}\eta^-_1-\nu_{bL}
\eta^0)]+H.c.
\label{331b}
\end{eqnarray}

\hspace{9.1cm}
\rule{-2mm}{0.1mm}\rule{8.7cm}{0.1mm}

\begin{multicols}{2}
\narrowtext

Hence we need a heavy fermion octet transforming as $({\bf1},{\bf 8},0)$ under
the 3-3-1 factors, 
\begin{eqnarray}
H_L=
\left (
\begin{array}{lcr}
N_1 & P^+_1 & P^+_2 \\
P^-_3 & N_2 & X^0_1 \\
P^-_4 & X^0_2 & N
\end{array}
\right )_L,
\label{octeto}
\end{eqnarray}
where $N=-N_1-N_2$.
We have the interaction
\begin{equation}
f_a[\overline{(\Psi_{aL})^c}\, H_L \eta +\eta^\dagger\,\overline{(H_L)^c}\,
\Psi_{aL}]+ H.c.,
\label{linkoct}
\end{equation}
omitting $SU(3)$ indices. There is also the bare mass term $M_H
\overline{(H_L)^c}H_L$. These interactions produce
a mass matrix for the neutral sector like that in Eq.~(\ref{masstriplet}), but
now with $M_{aL}=f_{a}\langle \eta \rangle$ and $M_\Delta\to M_H$. Two 
neutrinos remain massless and two gain masses. The tree level realization of
this mechanism can be seen in Fig.~\ref{fig1} with $N_R$ replaced by 
$(N_{1,2L})^c$ and $\langle \phi \rangle $ replaced by $\langle \eta \rangle $.
However, if we introduce three of such octets and identify, say $N_{1aL}$ with
$(\nu^c)_{aL},\,a=e,\mu,\tau$, we can generate a general see-saw mass matrix
like in Eq.~(\ref{D-Mmatrix}) since now $M_{Lab}=f_{ab}\langle \eta \rangle$.  
Notice that although this case is the analog of the case {\bf III)} in the
previous section, i.e., an hermitian, $Y=0$, fermion triplet; it is also a
realization of the case {\bf IV)} where
there is a non-Hermitian, $Y=2$, fermion triplet under $SU(2)_L\otimes U(1)_Y$.
 
As in the case of the leptonic triplet in the standard model, the mass matrix 
of the charged lepton sector is complicated. Let us assume, for simplicity, 
that the mass matrix of the standard charged leptons, generated at low energy 
in the basis of the 3-3-1 symmetries, is diagonal and let us call it $M_l$, 
and for the mass 
matrix of the charged leptons, $P_1$ and $P_2$, 
matrix $M_H$. With these assumptions and the terms in Eq.~(\ref{linkoct}), we
have the following $5 \times 5$ matrix in the basis $(l_L ,\, P_{1L},\, P_{2L}
)$: 
\begin{eqnarray}
\left (
\begin{array}{cc}
M_l & M_\eta  \\
0 & M_H  \\
\end{array}
\right ).
\label{chargedlep}
\end{eqnarray}
As in the case {\bf III)} in the previous section, the diagonalization of this
matrix is complicated unless we assume that $M_l$ is diagonal, in this case the
usual leptons decouple completely  from the heavy ones, being the mass matrix of
the former $M_l$ and that for the heaviest $M_H$. 

{\bf C}) Finally, we have the possibility that $\Psi^c_{aR}$ and $\Psi_{bL}$
form a symmetric sextet. In this case we have
\begin{equation}
{\cal L}^{eff}_{6s}=\frac{f_{ab}}{\Lambda}
(\overline{(\Psi^c)_{aR}}\,\Psi_{bL}\eta^\dagger\eta^\dagger)+H.c.,
\label{sextet}
\end{equation}
and the tree level realization is the introduction of the usual sextet
$({\bf6},0)$~\cite{331}. Notice that the triplet
$\epsilon_{ijk}\bar\Psi^c_j\Psi_k$ is not
realized since the respective triplet $\epsilon_{ilm}\eta_l\eta_m=0$. However, 
the symmetric sector can contribute for the masses of the charged leptons and
also generate the neutrino masses by loop effects. 

In this case the tree level realization requires a scalar sextet,
$S$:
\begin{eqnarray}
S=
\left (
\begin{array}{lcr}
s^{\prime} & \frac{s_1^{-}}{\sqrt{2}} & \frac{s_2^{+}}{\sqrt{2}} \\
\frac{s_1^{-}}{\sqrt{2}} & s_3^{--} & \frac{s}{\sqrt{2}} \\
\frac{s_2^{+}}{\sqrt{2}} & \frac{s}{\sqrt{2}} & s^{++}_4
\end{array}
\right )\sim({\bf 1},{\bf 6},0).
\label{sextet2}
\end{eqnarray}
The sextet $S$ is linked to the leptons by the interaction:
\begin{equation}
f_{ab}\bar \Psi^c_a S \Psi_b + H.c.
\label{sextetlink}
\end{equation}
When the neutral component $s^{\prime}$ of the sextet $S$ develops a 
VEV, the neutrino mass matrix get the following  form:
\begin{equation}
M_{\nu ab} = f_{ab}\langle s^{\prime} \rangle.
\label{neutmasssext}
\end{equation}

The next step is to show how $\langle s^{\prime} \rangle$ gets a tiny 
value. For this we need to develop the scalar potential of 
the model taking into account the explicit terms that violate the lepton 
number. Let us first write the scalar potential that conserves the lepton
number:  
\end{multicols}
\hspace{-0.5cm}
\rule{8.7cm}{0.1mm}\rule{0.1mm}{2mm}
\widetext

\begin{eqnarray}
V(\eta,\rho,\chi,S)&=&
\mu^2_\eta \eta^{\dagger}\eta+\mu^2_\rho 
\rho^{\dagger}\rho+\mu^2_\chi \chi^{\dagger}\chi+\mu^2_S 
Tr(S^{\dagger}S)+\lambda_1(\eta^{\dagger} \eta)^2 +\lambda_2 (\rho^{\dagger}\rho
)^2 +\lambda_3(\chi^{\dagger}\chi)^2 \nonumber \\ &+&
(\eta^{\dagger}\eta)\left( \lambda_4 
(\rho^{\dagger}\rho) + 
\lambda_5(\chi^{\dagger}\chi)\right)+\lambda_6(\rho^{\dagger}\rho)
(\chi^{\dagger}\chi)
+\lambda_7(\rho^{\dagger}\eta)(\eta^{\dagger}\rho)+\lambda_8
(\chi^{\dagger}
\eta)(\eta^{\dagger}\chi)+
\lambda_9(\rho^{\dagger} \chi)(
\chi^{\dagger}\rho) 
\nonumber \\ &+&
\lambda_{10}Tr(S^{\dagger}S)^2+\lambda_{11}\left( 
Tr(S^{\dagger}S) \right)^2 +\left(\lambda_{12}(\eta^{\dagger}\eta) + 
\lambda_{13}(\rho^{\dagger}\rho)\right)Tr(S^{\dagger}S)+ 
\lambda_{14}(\chi^{\dagger}\chi)Tr(S^{\dagger}S)\nonumber \\
&+& \left(\lambda_{15}\epsilon^{ijk}(\chi^{\dagger}S)_i \chi_j \eta_k + 
\lambda_{16}\epsilon^{ijk}(\rho^{\dagger}S)_i \rho_j \eta_k 
+ \lambda_{17}\epsilon^{ijk}\epsilon^{lmn}\eta_n \eta_k 
S_{li}S_{mj} + 
H.c. \right) + \lambda_{18}\chi^{\dagger}SS^{\dagger} \chi  \nonumber \\ &+& 
\lambda_{19}\eta^{\dagger}SS^{\dagger} \eta + 
\lambda_{20}\rho^{\dagger}SS^{\dagger} \rho.
\label{331potential}
\end{eqnarray}
\hspace{9.1cm}
\rule{-2mm}{0.1mm}\rule{8.7cm}{0.1mm}

\begin{multicols}{2}
\narrowtext

Here, for the sake of 
simplicity, we impose in the scalar potential the symmetry 
$\chi \rightarrow -\chi$ in order to avoid other trilinear terms  
besides the one that will generate the see-saw mechanism. 
There are also four terms permitted by the 3-3-1 gauge symmetry which violate
explicitly the lepton number, but for what concern us here we just will consider
only one of  them:
\begin{equation}
M_S \eta^T S^{\dagger} \eta.
\label{ptvlp}
\end{equation}

Adding this term to the potential in Eq.~(\ref{331potential}), we 
find the following stationary condition on the VEV of the scalar field 
$s^{\prime}$:

\end{multicols}
\hspace{-0.5cm}
\rule{8.7cm}{0.1mm}\rule{0.1mm}{2mm}
\widetext

\begin{equation}
\langle s^{\prime} \rangle 
\left[ \mu^2_S   +  \lambda_{10}\langle s \rangle^2 + 
( \frac{\lambda_{12}}{2} + \frac{\lambda_{19}}{2} )\langle \eta 
\rangle^2
+\frac{\lambda_{13}}{2}\langle \rho \rangle^2  + 
\frac{\lambda_{14}}{2} \langle 
\chi \rangle^2 + 
(\lambda_{10} + \frac{\lambda_{11}}{2}) \langle s^{\prime} \rangle^2 
\right]
+ M_S \langle \eta \rangle^2=0.
\label{mspp}
\end{equation}

\hspace{9.1cm}
\rule{-2mm}{0.1mm}\rule{8.7cm}{0.1mm}

\begin{multicols}{2}
\narrowtext

We will consider here the case when the charged leptons gain masses, not 
through a sextet but via the introduction of a heavy charged lepton~\cite{dma}. 
In this case the sextet can be very heavy.
For instance, we can assume $\mu^2_S<0$ and that $\vert\mu_S\vert= 
M_S \gg \langle \eta \rangle ,\langle\rho\rangle , \langle \chi \rangle $, 
so that we have from Eq.~(\ref{mspp}):
\begin{equation}
\langle s^{\prime} \rangle \simeq \frac{\langle \eta \rangle^2 }{M_S},
\label{seestrip}
\end{equation}
which gives the following expression to the neutrino mass matrix:
\begin{equation}
M_{\nu ab}= f_{ab}\frac{\langle \eta \rangle^2 }{M_S}.
\label{ss331s}
\end{equation}
This is a see-saw mass relation and its realization is equal to that in 
Fig.~\ref{fig2} with $\langle \xi \rangle$ replaced by 
$\langle s^{\prime} \rangle$ and 
$\langle \phi \rangle$ replaced by $\langle \eta \rangle$. It is the equivalent
to the type II see-saw mechanism  discussed in {\bf II}) in the last section 
and in Ref.~\cite{typeII} where it was assumed that the sextet also
gives mass to the charged leptons but the dominant energy scale is
$\langle\chi\rangle$.

A sextet like that in Eq.~(\ref{sextet2}) disposes of a second
neutral component, $s$, which may contribute to the charged lepton masses when 
developing a VEV. However it will not give a large contribution to the charged
lepton masses as we can see in the following. From its stationary condition 
we get the constraint equation:

\end{multicols}
\hspace{-0.5cm}
\rule{8.7cm}{0.1mm}\rule{0.1mm}{2mm}
\widetext

\begin{eqnarray}
&&\langle s \rangle \left[ \mu^2_S   + \lambda_{10} \langle s^{\prime} 
\rangle^2 
+ 
(\frac{\lambda_{12}}{2} -\lambda_{17} )\langle \eta \rangle^2 +( 
\frac{\lambda_{13}}{2}  + 
\frac{ \lambda_{20} 
}{4} )\langle \rho \rangle^2 + (\frac{ 
\lambda_{14} 
}{2}   + \frac{ \lambda_{18} }{ 4 })\langle \chi \rangle^2 + 
(\frac{\lambda_{11}}{2} + 
\lambda_{10})\langle s\rangle^2\right]\nonumber \\
&&+ \frac{ \lambda_{15} }{ 2\sqrt{2}}\langle \eta \rangle \langle \chi 
\rangle^2 
- \frac{ \lambda_{16} }{ 2\sqrt{2} }\langle \eta \rangle \langle \rho 
\rangle^2 
= 0,
\label{mcs2e} 
\end{eqnarray}
\hspace{9.1cm}
\rule{-2mm}{0.1mm}\rule{8.7cm}{0.1mm}
\begin{multicols}{2}
\narrowtext
which gives the following expression to its vacuum:
\begin{equation}
\langle s \rangle \simeq \frac{\langle \eta \rangle \langle \chi 
\rangle^2 }{\mu^2_S}.
\label{vacsig2}
\end{equation}
It implies $\langle s \rangle \ll \langle s^{\prime} \rangle$ which 
turn insignificant its contribution to the charged lepton masses in relation to
other sources in low energy. Notwithstanding, in the present case the charged
leptons got mass through the mixing with a heavy charged lepton~\cite{lepmass1}.

\section{conclusions}
\label{sec:con}

We have analyzed  the realization at the tree level of an 
effective dimension-five operator that generates  see-saw masses to the 
neutrinos and have added a new tree level implementation of the see-saw
mechanism in the context of the standard electroweak model.
From our analysis we can say that all the ways of  realizing  such
operator in the standard model can be easily implemented in the 3-3-1 model
as well. In particular the case {\bf IV)} of Sec.~\ref{sec:sm} is realized in
the context of the 3-3-1 model when a fermion octet is added as described in
the case {\bf B}) in Sec.~\ref{sec:331}.

\begin{figure}
\begin{center}
\includegraphics[width=5cm]{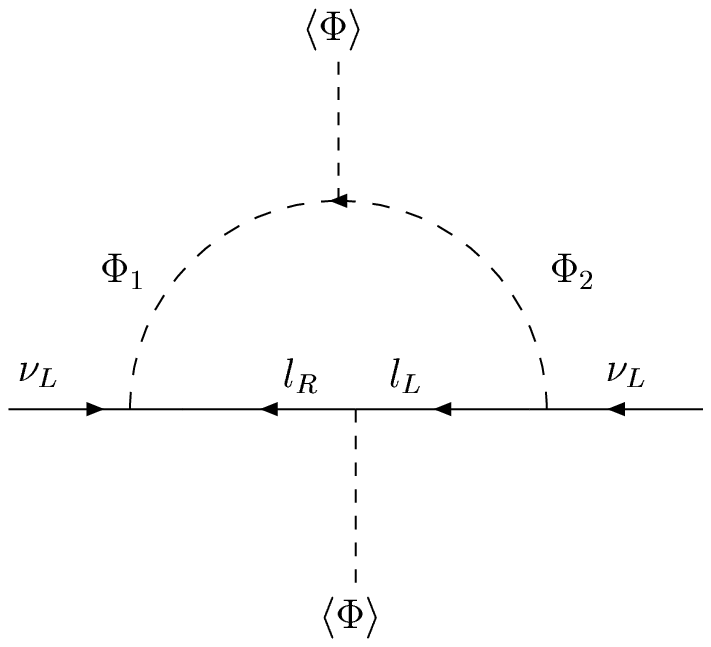}
\end{center}
\caption{
One loop realization of the effective 
dimension-five operator  
}
\label{fig3}
\end{figure}

\begin{figure}
\begin{center}
\includegraphics[width=5cm]{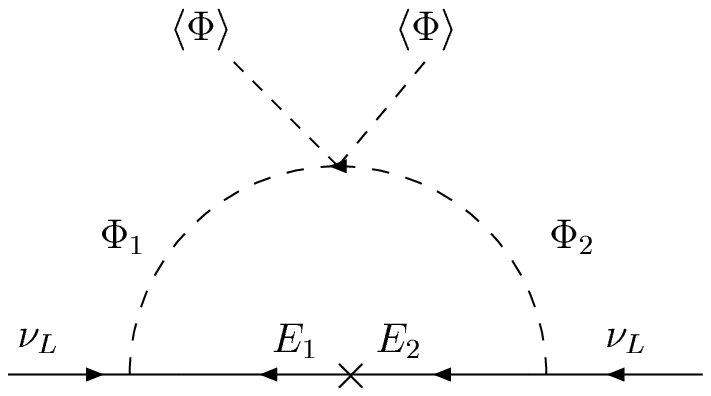}
\end{center}
\caption{
Another one loop realization of the 
effective dimension-five operator. 
}
\label{fig4}
\end{figure}

Finally, we would like to mention that at the one loop level the effective
operator of the sort given in Eq.~(\ref{e1}) is also easily implemented. Two
simple cases are shown in Figs.~\ref{fig3} and \ref{fig4}, 
but they imply an extension of the 3-3-1 model by adding a pair of exotic
lepton singlets. Only the second model is favored by phenomenology. Also we
should stress that, except in case {\bf B})  the neutrinos and
charged lepton masses share a common set of parameters which leave such
scenarios interesting.   

\acknowledgments 
This work was supported by Funda\c{c}\~ao de Amparo \`a Pesquisa
do Estado de S\~ao Paulo (FAPESP), Conselho Nacional de 
Ci\^encia e Tecnologia (CNPq) and by Programa de Apoio a
N\'ucleos de Excel\^encia (PRONEX).

\end{multicols}


\vspace{5mm}
\widetext

\end{document}